\documentclass[runningheads,a4paper]{llncs}

\usepackage{amssymb}
\usepackage{graphicx}
\usepackage{url}
\usepackage{apacite}
\usepackage{amsmath}
\usepackage{hyperref}
\newcommand{\keywords}[1]{\par\addvspace\baselineskip
\noindent\keywordname\enspace\ignorespaces#1}
\usepackage{booktabs}
\usepackage{url}

\newcommand{\tensiledirection}{tensile\_strain\_direction}
\newcommand{\diameterdirection}{cloud\_diameter\_direction}
\newcommand{\tensilelevel}{tensile\_strain\_level}
\newcommand{\diameterlevel}{cloud\_diameter\_level}

\usepackage{natbib}

\begin{document}

\mainmatter  

\title{A variational autoencoder for music generation controlled by tonal tension}


%
%
\author{Rui Guo\inst{1}\and Ivor Simpson\inst{1}\and Thor Magnusson\inst{1}\and Chris Kiefer\inst{1}\and Dorien Herremans\inst{2}}
\authorrunning{Rui Guo et al.}
\titlerunning{A variational autoencoder for music generation controlled by tonal tension}
\institute{University of Sussex \and {Singapore University of Technology and Design}\\}
%


%
%

\maketitle

\begin{abstract}
Many of the music generation systems based on neural networks are fully autonomous and do not offer control over the generation process. In this research, we present a controllable music generation system in terms of tonal tension. We incorporate two tonal tension measures based on the Spiral Array Tension theory into a variational autoencoder model. This allows us to control the direction of the tonal tension throughout the generated piece, as well as the overall level of tonal tension. Given a seed musical fragment, stemming from either the user input or from directly sampling from the latent space, the model can generate variations of this original seed fragment with altered tonal tension. This altered music still resembles the seed music rhythmically, but the pitch of the notes are changed to match the desired tonal tension as conditioned by the user. 
\keywords{music generation, generative model, variational autoencoder, tonal tension }
\end{abstract}

\section{Introduction}

Automatic music generation systems date back centuries. For instance, one famous example dates back to the 18th century, when people played the musical dice game to generate new music using a different (probabilistic) combination of musical bars \citep{herremansFunctionalTaxonomyMusic2017}. Deep learning has caused a steep increase in popularity in this ancient field \citep{Huang2019MusicTG}. However, giving the user high level control over the music that is being generated is still to be explored. When we allow for controllability, the resulting system can be used for narrative purposes such as film and game music. In this paper, we offer such a controllable system, with a focus on an important aspect of music: tension.

\section{Related Work}
A lot of work has been done on music generation systems in the last few decades. We refer the reader to \citet{Briot2020FromAN,herremansFunctionalTaxonomyMusic2017} for a more complete overview. When we are able to control certain aspects of the generated music, we open up opportunities for co-creation between artist and machine. When considering high level controls for music generation, one particular feature of interest is tonal tension, which is strongly related to emotion~\citep{marcocostaPERCEIVEDTENSIONMOVEMENT,meyerEmotionMeaningMusic1956}. Controlling the tension gives us a way to control part of the affect in generated music. Tension in music often endows a feeling of cohesion, e.g. one might see two music phrases with the first one exhibiting a rise in tension followed by a phrase with a release in tension.

When it comes to generating music conditioned by tension, the existing research is limited. In a system by \citet{farbood_2007}, a user-inputted harmony line guides the generated chord progressions. In another system, MorpheuS~\citep{herremansMorpheuSGeneratingStructured2017} uses a variable neighborhood search optimization algorithm to generate music with specified tonal tension shapes.  MorpheuS morphs the pitches of an existing input piece so that the resulting piece matches a given tonal tension shape, while preserving the musical structure (i.e. repeated patterns). \citet{williams2017affective} proposed a neural network to adjust five musical features so as to generate music with the widest spread of stimuli in the valence-arousal space. Other related research uses long-short term memory (LSTM) networks to generate a tension profile first and uses this to condition the music generation~\citep{verstraelenGeneratingMusicCoherent}. The author reports the generated tension profile does not work as well as the template tension input. In \citet{tan2020fadernets}, Music FaderNets are used to change the amount of arousal in short, generated, musical fragments. In the current research, we aim to further improve the state-of-the-art by proposing to integrate two tonal tension measures into a variational autoencoder model to change the tonal tension of a seed fragment, while keeping the seed music rhythm mostly unchanged. 

\section{Method}

\subsection{Tension Measures}\label{sec:tension}

Musical tension may come from a variety of sources such as tempo, rhythm and timbre~\citep{farbood2006quantitative, farbood2017contribution, herremans2017imma}. In this work, we focus on tonal tension, measured with a model based on the Spiral Array theory~\citep{chewSpiralArrayAlgorithm2002,chew2014mathematical}. The Spiral Array is a three-dimensional geometrical model which represents the tonal space. It consists of three spirals, one that represents the position of all pitches, one for chords, and one for keys. Within this geometrical space, a closer tonal distance results in a closer geometrical distance. 

\citet{herremansTensionRibbonsQuantifying} developed a model for tonal tension based on the Spiral Array. This model captures three aspects of tonal tension: cloud diameter, tensile strain, and cloud momentum. In this paper, we focus on the first two.

The \textbf{cloud diameter} captures how ``tonally close'' the pitches are in the tonal space, i.e. tension in terms of dissonance. To calculate this, we split our musical piece into windows (or clouds of notes). For each window, the cloud diameter represents the largest distance between notes of the cloud. A moving average window of one quarter note is used to smooth the diameter curve and prevent abrupt changes from one 16th note to another. The same applies for the tensile strain defined below.

We also calculate the \textbf{tensile strain} for each cloud of notes. This is defined by the distance between the geometric gravity point of all the pitches in the cloud (i.e. the center of effect). and the geometric position of the key of the piece.


\subsection{Training data} 

A total of 7,289 MIDI files were selected from the LMD-matched dataset~\citep{raffel2016learning} with the tag ``pop'' as calculated by \citep{magenta}. The Midi-Miner tool \citep{guoMidiMinerPython2019} was used to extract both the melody and bass tracks from the MIDI files, and subsequently calculate two tonal tension measures described in the previous section. After that step, a total of 3,457 files are valid (i.e. contain both melody and bass track) to use as training data. The melody and bass track are both monophonic after preprocessing by Midi-Miner. A bass track is included to make the harmony progression clearer and the tension more perceptible. Each song was divided into several four-bar long fragments with both a melody and bass track, resulting in 44,900 four-bar long fragments. All of the songs were first transposed to C major or A minor (depending on their mode) and the key position $key_{pos}$ of all the songs is set as C major to calculate the tensile strain measure. It should be noted that the A minor $key_{pos}$ can also be used if the key of the song is given to the model, which is not the case here.

The input for the variational autoencoder (VAE) model is a $64\times89$ piano roll,  whereby 89 is the feature dimension and 64 is the time dimension. One time step accounts for a 16th note, and 64 time steps equal four bars length in 4/4 meter. The 89 dimensions are comprised of four feature sets: melody pitches, melody onsets, bass pitches and bass onsets. The melody's pitches are represented as a 74-dim one-hot vector in the MIDI pitch range of $[24,96]$. Pitches outside of that range will be omitted so as to focus on the most frequently occurring pitches. The last dimension of melody pitch vector marks a rest note. The melody's rhythm is also represented by a one-hot vector in which 1 represents a new note at that time step and 0 no new note. The bass' pitch is represented as a 13-dim one-hot vector which maps to 12 pitch classes plus a rest note (the last dimension). The bass' rhythm representation is similar to the melody's rhythm representation.


\subsection{Model Details}

\begin{figure}[htb]
 \centerline{\framebox{
 \includegraphics[width=0.8\columnwidth]{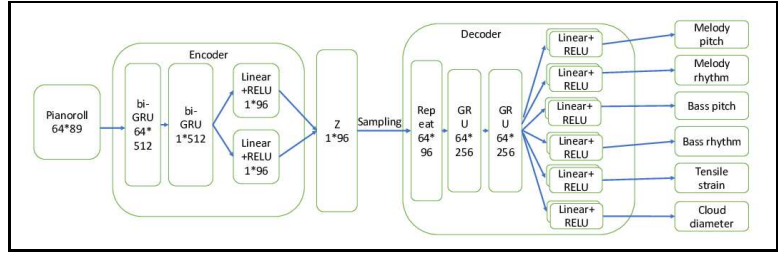}}}
 \caption{Proposed model architecture. The input is a 64*89 piano roll with melody and bass tracks, and the output includes the reconstruction of the piano roll and the tension measures.}
 \label{fig:model}
\end{figure}

The model is shown in Figure~\ref{fig:model}.  A VAE~\citep{kingma2013autoencoding} encoder parameterised by $\phi$ is used to map the input piano roll $x$ to a latent space $z$. The decoder parameterised by $\theta$ maps $z$ to 6 separate outputs $[y_1..y_6]$. $[y1..y4]$ is the reconstruction of the input $x$ which includes four feature sets. The original VAE loss is the reconstruction loss minus the KL divergence of the unit Gaussian prior $p(z)$  and the posterior distribution $q_{\phi}(z|x)$. $\beta$ is used to change the balance of reconstruction loss and the Kullback–Leibler divergence loss.

\begin{align}
\mathcal{L}_{vae}  & = L_{rec}(\theta,\phi,x) - 
\beta D_{KL}(q_{\phi}(z|x)||p(z))  \nonumber \\  & = \mathbb{E}_{q_{\phi}(z|x)} (\log{p_{\theta}(x|z)})  - \beta D_{KL}(q_{\phi}(z|x)||p(z)) 
\end{align}

In addition to the original VAE loss, the tension loss based on the predicted tension $y_5$ and $y_6$ are added, which are the tensile strain output and cloud diameter output respectively. $tensile\_strain(x)$ and $diameter(x)$ are tension measures calculated from the input $x$ by the spiral array theory. MSE is the mean square error loss. The tension loss is defined by:

\begin{equation}
\mathcal{L}_{tension}   = MSE(tensile\_strain(x), y_5) + MSE(diameter(x), y_6)
\end{equation}

The total loss for the model is hence  given as: 

\begin{equation}
\mathcal{L}_{total}   =  \mathcal{L}_{vae} +   \mathcal{L}_{tension}
\end{equation}

A two layer gated recurrent unit (GRU)\citep{cho2014properties} with 256 nodes is used in both the encoder and decoder. In the decoder, the latent variable is repeated for 64 timesteps to feed into the two GRU layers then followed by two dense layers for each of the six outputs. $\beta$ is set to 0.006 with KL annealing~\citep{bowman2015generating} of an increase of 5e-7 for each batch until 0.006. The training, validation and test dataset split ratio is 0.8, 0.1, 0.1 respectively. The Adam optimiser is used with a start learning rate of 0.001. The model's test set loss is used for early stopping to mitigate overfitting. 

\section{Experiment and evaluation}

In this section, we first compare the loss of the proposed model with that of models that do not output tension measures. Then we explore the latent space and identify four latent vectors which allow us to control the tension direction or overall tension level of the generated music. The effect of those tension vectors is validated and finally, possible applications of those latent vectors are discussed.

\subsection{Model loss comparison}

We compare the proposed model with several variants. The baseline model simply reconstructs the input piano roll. Other variants of the model jointly learn to generate tension values by incorporating this information during the model training. Each model was run five times and the result for the best run was used as shown in Table~\ref{tab:model_result}. Please note that additional loss terms were added for the cases in which tension was present. The KL loss is weighted by $\beta=0.006$ in the total loss. When learning to predict tensile strain, the model bass pitch loss decreased. When predicting both the tension measures, the rhythm loss does not change much compared to the baseline model.

\begin{table}[]
\begin{center}
\scriptsize
\begin{tabular}{lccccccccc}
\toprule
Model           & Total loss & Melody pitch & Melody rhythm & Bass pitch & Bass rhythm & Tensile & Diameter &  KL  \\ \midrule
Baseline            & 2.025 & 0.5335        & 0.1723         & 0.4120      & 0.1463       & NA      & NA         & 128 \\ 
Added ts & 2.0816 & 0.5457        & 0.1737         & 0.4080      & 0.1533       & 0.049   & NA         & 126 \\ 
Added cd & 2.1942 & 0.5304        & 0.2998         & 0.41      & 0.1551       & NA      & 0.1244       & 129 \\ 
Proposed  & 2.2334 & 0.5405        & 0.1967         & 0.3994      & 0.1520       & 0.0454   & 0.1197       & 129 \\ \bottomrule
\end{tabular}
\end{center}
 \caption{Model loss with different conditional outputs. The baseline model only reconstructs the input piano roll, the other models generate not only the reconstructed input, but also additional features such as tensile strain (ts) and cloud diameter (cd). The proposed model learns to generate two tension measures along with reconstructing the input piano roll. }
\label{tab:model_result}
\end{table}

\subsection{Identifying tension feature vectors}

We are interested in using the latent space to manipulate the tension of the generated music. To understand how certain aspects of tonal tension are captured by the latent variables in our model, we have identified two types of latent feature vectors that we can use to change the music tension learned in the model to create specific manipulations of the generated music tension. One type allows us to control the evolution of the tension direction from the beginning of the 4 bar fragment to the end. The other controls the overall tension level throughout the 4 bars. We first assign tension class labels to our music fragments (upwards/downwards tensile strain/cloud diameter and high/low tensile strain/cloud diameter). To assign the first pair of labels, the correlation between the tension throughout the fragment and a straight line that goes through (0,0) to (1,1) with slope of 1 is calculated. For the second pair of labels, the 2-norm of difference of the tension value and a threshold value is calculated. Different thresholds for the correlation and 2-norm are used to select around 1,000 samples for each of the class labels. We adopt an approach similar to \citep{hou2017deep} to identify the feature vectors. Four latent feature vectors are identified: $\tensiledirection$, $\tensilelevel$, $\diameterdirection$, and $\diameterlevel$.


\subsection{Validation of tension feature vectors}
In this subsection, we evaluate the effectiveness of the four feature vectors identified in the previous section and how they interact with each other.  

\subsubsection{Influence of each tension feature vector}

We random sampled 10,000 $z$ from the latent space to conduct the following experiments. First we add the four different tension feature vectors multiplied by a scaling factor to the original sampled $z$ and then decoded from this new latent vector. We expect that: a) the tension of the generated music should change according to the definition of the chosen tension feature; b) the generated music should keep the original rhythm and not much changed compared with the original music as tonal tension is not influenced by rhythm.

\begin{figure}[htb]
 \centerline{\framebox{ 
 \includegraphics[width=0.9\columnwidth]{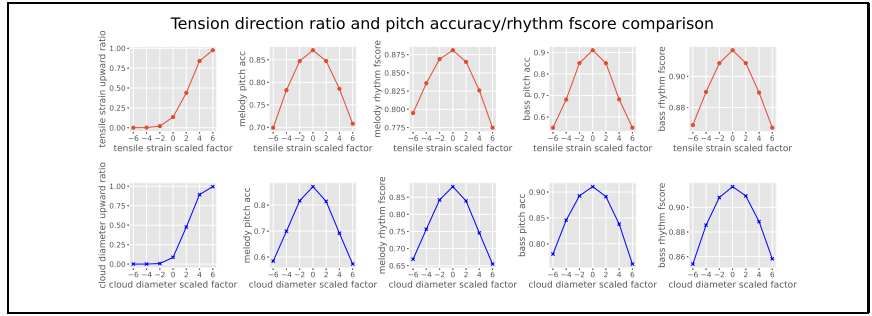}}}
 \caption{Tension direction ratio and pitch accuracy/rhythm F-score comparison with scaled direction feature vectors. The pitch accuracy change is higher than the rhythm F-score change, and the melody rhythm is more affected by than the bass rhythm.}
 \label{fig:tension_direction}
\end{figure}

Figure~\ref{fig:tension_direction} shows the influence of adding different scaled tension direction feature vector to the latent space of each sample from the selected dataset. The tension upward ratio is calculated by the number of generated pieces with with upward tension divided by the total number of generated pieces. From the most left figures, we can see that the upward ratio of those tension measures increases with a larger scaled feature vector added to the original latent vector. These results confirm that adding this vector to the latent space has a significant influence on both the tensile strain and cloud diameter.  In the middle graphs, we see that the change in melody rhythm by adding the $\tensiledirection$ feature vector is small, while the $\diameterdirection$ change the melody rhythm is slightly higher. The bass rhythm F-score does not change much compared to the bass pitch accuracy and even less than the melody rhythm F-score.

From these results, we postulate that $\tensiledirection$ changes the harmony progression and keeps the rhythm, while the $\diameterdirection$ adds/reduces notes with high tension to the melody resulting in a rhythm change.  This was equally confirmed for the $\tensilelevel$ and $\diameterlevel$ in Figure~\ref{fig:tension_level} in Appendix.

\subsubsection{Interaction among feature vectors}
\begin{figure}[htb]
 \centerline{\framebox{
 \includegraphics[width=0.8\columnwidth]{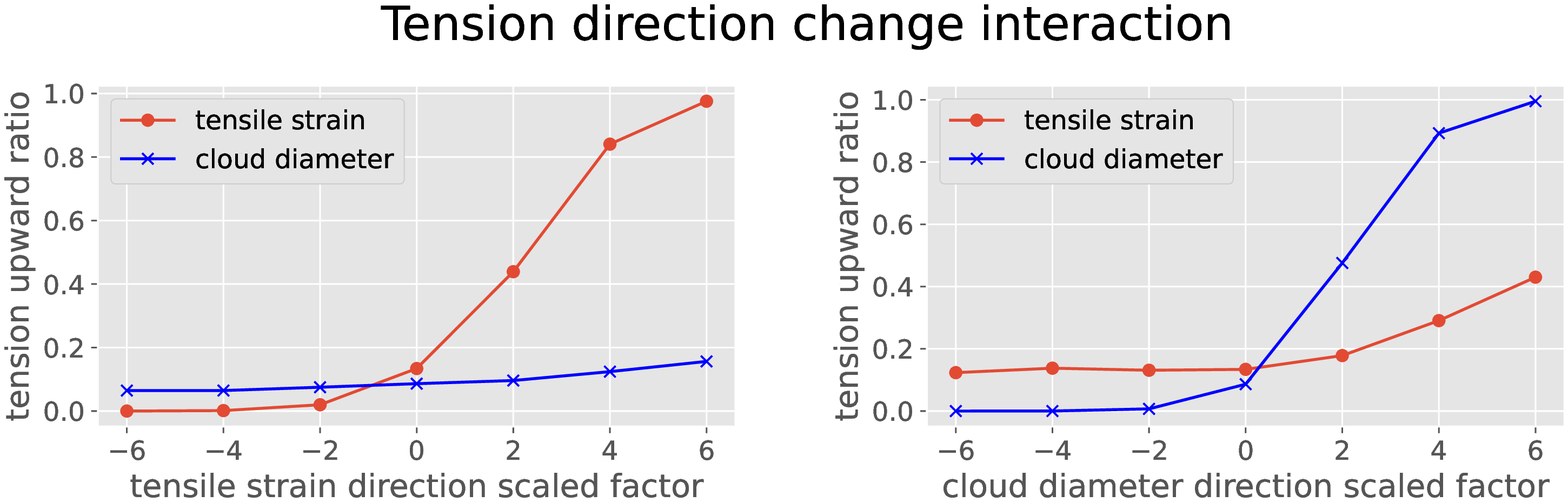}}}
 \caption{Tension direction changes by varying the applied scaled $\tensiledirection$ or $\diameterdirection$. The upward ratio is calculated by the number of the output with tension upward shape divided by the total number of samples.}
 \label{fig:tension_direction_interaction}
\end{figure}

We also explored the interactions between different tension feature vectors so as to find out how much they can influence the tension output controlled by the other feature vector, as well as the function they play in the generation of music. In particular, the interaction between $\tensiledirection$ and $\diameterdirection$ is explored in order to examine if applying the $\tensiledirection$ vector can change the cloud diameter prediction and vice versa. In Figure~\ref{fig:tension_direction_interaction}, the $\tensiledirection$ or $\diameterdirection$ with varying scaling factors are added to the latent space $z$ and the tensile strain/cloud diameter upward ratio is shown. A change of the scaling factor for $\tensiledirection$ does not affect the cloud diameter upward ratio a lot, however, the $\diameterdirection$ changes the tensile strain upward ratio more. This shows that applying a change in tensile strain does not necessarily change the cloud diameter of the output, and a change in the cloud diameter of the output is more likely to change the tensile strain, which can be explained by their definition (Section \ref{sec:tension}). A similar analysis of the tension level feature vector can be found in Figure~\ref{fig:tension_level_interaction} in Appendix.

\begin{figure}[htb]
 \centerline{\framebox{
 \includegraphics[width=0.7\columnwidth]{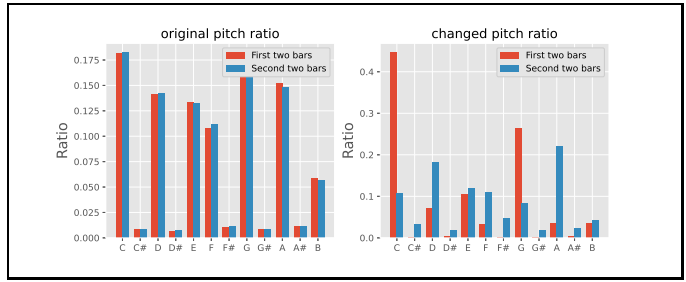}}}
 \caption{The pitch distribution of the original music and music with upward tensile strain direction. The number of occurrences of both notes C and G drops and of notes A, E, D rises significantly in the second two bars of the tensile strain upward music compared to the original music.}
 \label{fig:pitch}
\end{figure}

We also performed a number of experiments to compare the pitch distribution of the original music as well as the changed versions. In Figure \ref{fig:pitch}, the pitch distribution of the original music and the music with added $6*\tensiledirection$ is compared. The frequency of note C and G decreases and note A, E, D occurs in a much higher proportion in the music with tensile strain upward shape.

In addition to the feature vector identified, our proposed method can change the tensile strain or cloud diameter into an arbitrary shape if a tension vector with a specific shape defined and identified. To validate this, a $\diagup\diagdown$ shape tensile strain attribute vector was found. Adding this scaled vector to the latent space allows us to generate music with a tensile strain shaped in the form of $\diagdown\diagup$ or $\diagup\diagdown$.

\subsubsection{Musical pieces}
A demonstration of modifying the music tension by adding the $\tensiledirection$ and $\diameterlevel$ to the latent space is shown in Figure \ref{fig:fivestaff} in the Appendix. Although this model can only output 4 bar long music, it can be used to create different variations within a larger musical piece. For instance, we can generate the first 8 bars of music by adding the scaled $\tensiledirection$ vector to the latent space of given input/sampled music, and then generate another 8 bars by first adding randomly scaled \newline $\diameterlevel$ to the first generated 8 bars' latent space vector. In this way, a variation of arbitrary length can be generated, which still sounds coherent to the other fragments as the same seed is used. The reader is invited to listen to some generated fragments and their tension figures at\newline \url{https://ruiguo-bio.github.io/tension_vae.github.io} and explore our online colab notebook \url{https://github.com/ruiguo-bio/colab_tension_vae}.

\section{Conclusion}

We propose a generative VAE model to control the tonal tension in generated music. Through experiments, we have shown that our system can modify the tonal tension in generated music based on either a existing or a newly sampled seed fragment. We have successfully identified ``tension feature vectors'', which can achieve different transformations of the output music by adding the scaled tension feature vector to the latent space. In our experiments, we show the following:  1) By adding a scaled tension feature vector to the latest space variables of the seed music, we can generate music with increasing tension, both in terms of tensile strain as well as cloud diameter (depending which tension feature vector is used); 2) Using a similar method we can also increase or decrease the overall increase or decreased tension level 3) Additional tension feature vectors can be found that can realise different shapes of tension in the generated music. 4) The bass pitch loss becomes less by incorporating the tension strain feature into the model.


In the future work, we would like to explore the inclusion of more musical tension factors. Besides the two tension measures used in this research, musical tension is also related to rhythm, timbre and many other factors. Meaningful musical properties \citep{wangExploringInherentProperties2020} can be fed into the system so as to further control the music generation. An interactive tool will be helpful to compose music by drawing tension. By using this tool to generate different variations to form a complete piece, we will examine how this affects the long-term structure and coherence of the generated music.

\subsubsection*{Acknowledgements}

This project was supported by the China Scholarship Council and MOE T2 grant no. MOE2018-T2-2-161.

\bibliographystyle{apacite}
\scriptsize
\bibliography{csmc2020}

\newpage

\section{Appendix}\label{sec:appendix}

\begin{figure}[htb]
 \centerline{\framebox{ 
 \includegraphics[width=0.9\columnwidth]{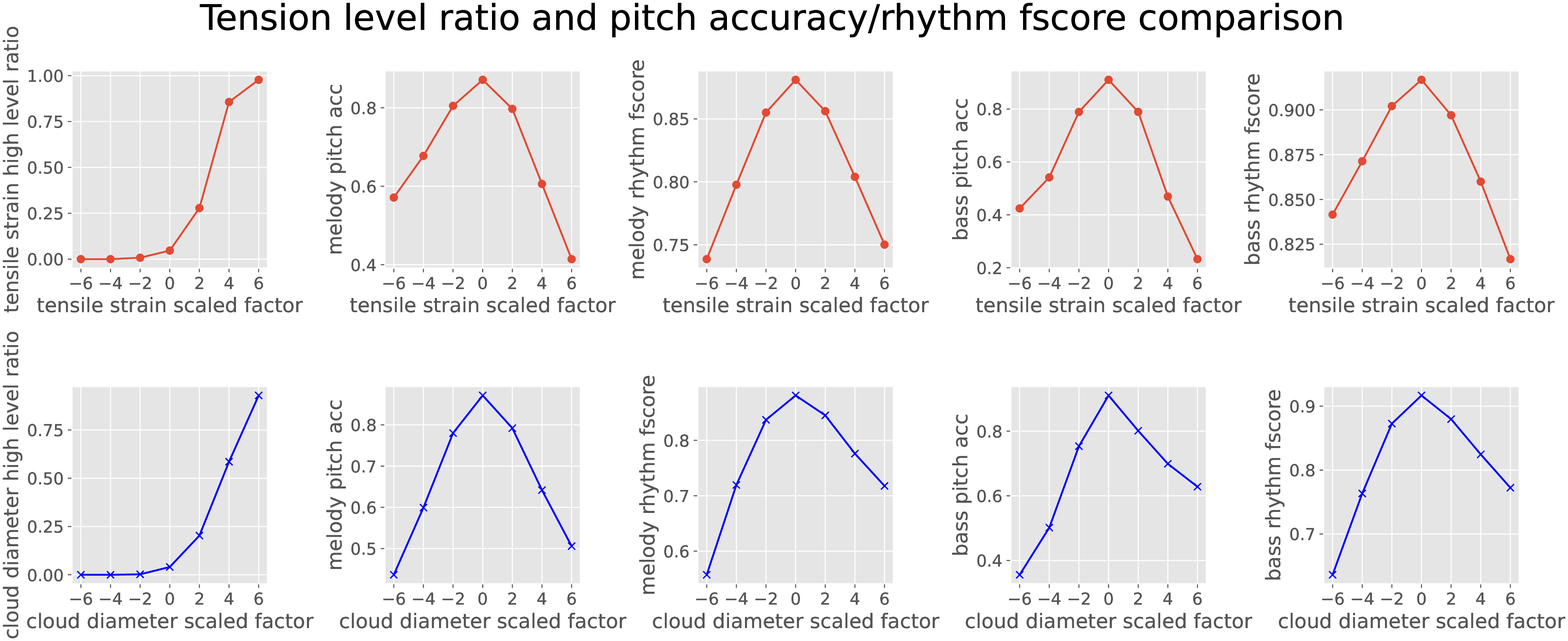}}}
 \caption{Tension level and reconstruction accuracy change comparison by adding different scaled $\tensilelevel$ or $\diameterlevel$ to the latent space of 10,000 random samples of the dataset. The high level ratio of tensile strain and diameter correlated with a larger scaling factor. The rhythm F-score is much less affected than the pitch accuracy vector, and $\diameterlevel$ changes rhythm more than $\tensilelevel$.}
 \label{fig:tension_level}
\end{figure}

\begin{figure}[htb]
 \centerline{\framebox{
 \includegraphics[width=1\columnwidth]{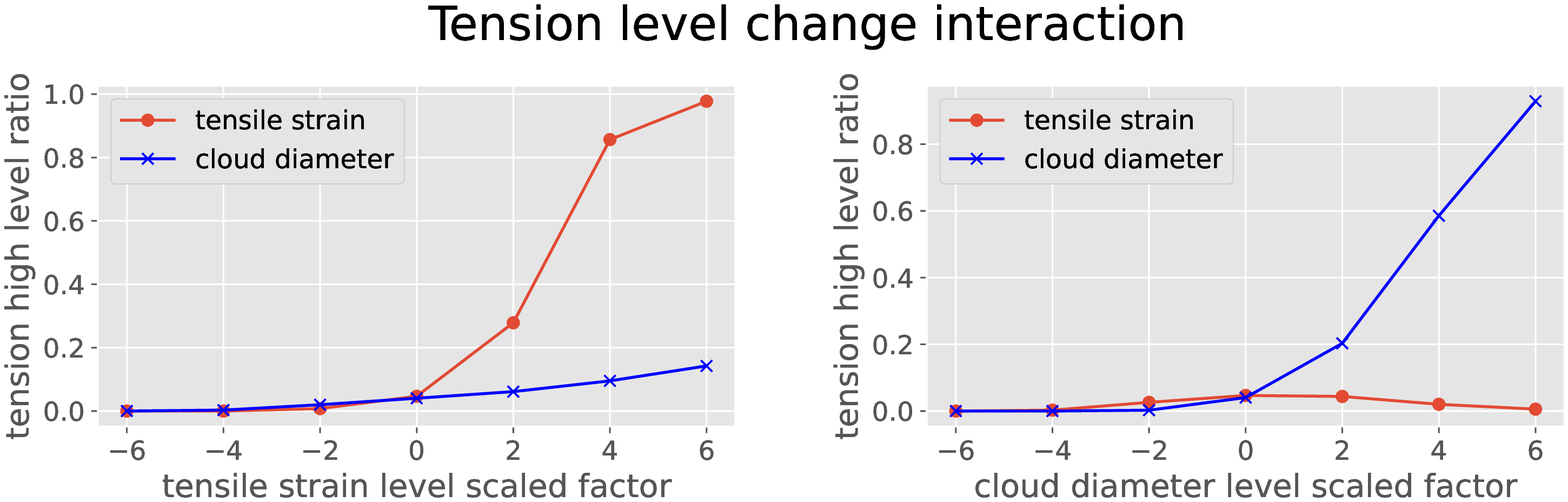}}}
 \caption{Tension level changes by adding only one of the scaled $\tensilelevel$ or $\diameterlevel$ to the latent space. The high level ratio is calculated by the number of the output with tension high level divided by the total number of samples. The scaled $\diameterlevel$ changes the tensile strain level more than the the reverse, which can be explained by their definition.}
 \label{fig:tension_level_interaction}
\end{figure}

\begin{figure}[htb]
 \centerline{\framebox{
 \includegraphics[width=1\columnwidth]{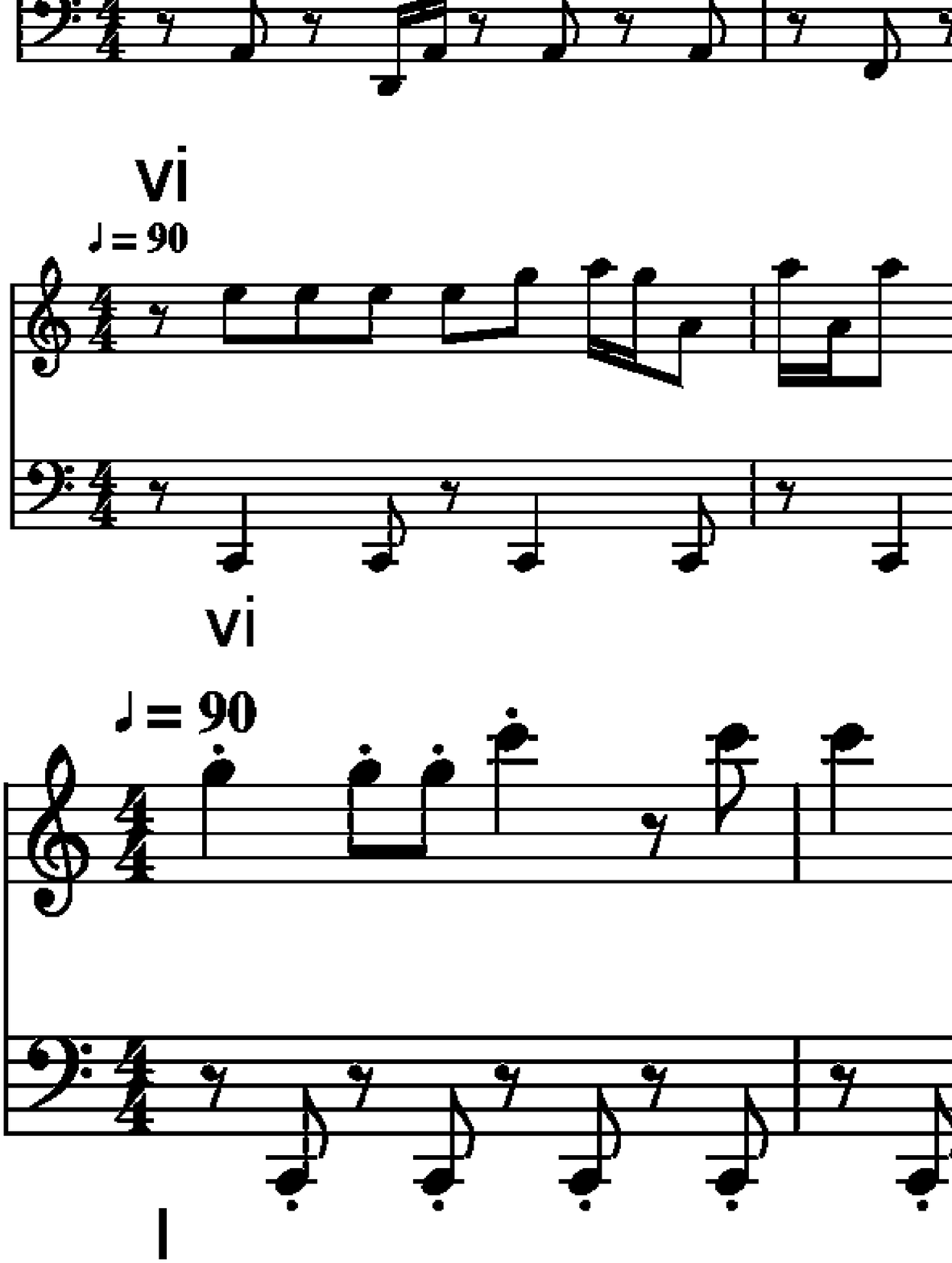}}}
 \caption{Music variations generated by changing $\tensiledirection$ and $\diameterlevel$ and their tension predictions. The top staff is generated music by sampling the latent space, the second and third staff is the music changed by adding/subtracting $6*\tensiledirection$ vector to the latent space of the original sample, and the fourth and fifth staff is music changed by adding/subtracting $3*\diameterlevel$ vector to the latent space of the original sample. The corresponding tension for those five staff is right side of the figure. The music generated by adding $6*\tensiledirection$ to the latent space changed its beginning harmony to C major and the harmony in the last two bar changes to A minor, and reverse is for the negative scaled $-6*\tensiledirection$ music. The cloud diameter level up version has more notes increasing the tonal tension of the whole piece, and the diameter level down version has less tension by reducing the notes with higher tension.  This shows the change of $\diameterlevel$ add/reduce the density of notes to increase/decrease the cloud diameter, resulting in the change of rhythm. This phenomenon is the same when we change the $\diameterdirection$ before. Although the cloud diameter definition does not contain rhythm information explicitly, more or less notes with higher tension will change rhythm as result.}
 \label{fig:fivestaff}
\end{figure}


\end{document}